\documentclass{aa}%

\usepackage[varg]{txfonts}
\usepackage{amsmath}
\usepackage{amssymb}

\begin{document}%

\title{Background subtraction and transient timing with Bayesian Blocks}
\author{H.~Worpel \and A.~D.~Schwope}
\institute{Leibniz-Institut f\"ur Astrophysik Potsdam (AIP), An der Sternwarte 16, 14482 Potsdam, Germany}

\date{March 16, 2015 (accepted)}

\abstract {}
   {To incorporate background subtraction into the Bayesian Blocks algorithm so that transient events can be timed accurately and precisely even in the presence of a substantial, rapidly variable, background.}
   {We developed several modifications to the algorithm and tested them on a simulated \emph{XMM-Newton} observation of a bursting and eclipsing object. }
   {We found that bursts can be found to good precision for almost all background subtraction methods, but eclipse ingresses and egresses present problems for most methods. We found one method that recovered these events with precision comparable to the interval between individual photons, in which both source and background region photons are combined into a single list and weighted according to the exposure area. We have also found that adjusting the Bayesian Blocks change points nearer to blocks with higher count rate removes a systematic bias towards blocks of low count rate.}
   {}

   
\maketitle%
\section{INTRODUCTION }
\label{sec:intro}

The detection and precise timing of transient events is of importance in all areas of astrophysics. In some applications, for instance X-ray astronomy, the desired timing accuracy is comparable to the interval between the arrival times of individual photons. In this situation it is necessary to identify unambiguously which photons mark the end of the pre-transient phase and the beginning of the post-transient phase and, if possible, estimate without systematic bias where between those photons the change occurs.

One method that has attracted recent interest is the Bayesian Blocks algorithm \citep{Scargle1998,ScargleEtAl2013}. This algorithm may take as its input a list of photon arrival times, such as those produced by X-ray telescopes. The algorithm can also handle other data modes, but in this work we consider exclusively time tagged event data. The observation is then divided into \emph{cells}, intervals containing a single photon, and periods of time of nominally constant count rate are produced by finding an optimum number and placement of \emph{blocks} of consecutive cells. Given a prior on the number of blocks, the algorithm finds an objectively optimal set of blocks describing the observation, and within each block the arrival times of the photons are consistent with a Poisson process with constant rate. The Bayesian Blocks algorithm is ideal for finding and accurately timing transients, as a \emph{change point} between blocks, typically placed half-way between the last event of one block and the first event of the next, will be found if and only if the rate of arrival of photons detectably changes. Transient timing can therefore, in principle, be performed with a precision comparable to the interval between individual photons, and clearly no better precision is possible. This property makes the Bayesian Blocks algorithm an attractive candidate for transient timing.

Other binning methods exist which, similarly to the Bayesian Blocks method, place bin edges at locations only where the data justifies it (e.g., \citealt{Knuth2006, Belanger2013}), but we have not considered them in this paper. See also \cite{Burgess2014} for a comparison of binning methods, in the context of gamma ray burst timing. 

The motivation for this paper is accurate and precise eclipse timings of cataclysmic variables (CVs) and low-mass X-ray binaries (LMXB) using data from the \emph{XMM-Newton} X-ray observatory. Much of the paper is presented with these applications in mind, but the methods and conclusions developed are not specific to this astronomical field and are not even limited to astronomy. We consider two effects that may affect accurate timing measurements using the Bayesian Blocks algorithm.

Firstly, is the location of the cell edges between photons optimal? While most applications put the cell edges half way between two photons (e.g., \citealt{Scargle1998, ScargleEtAl2013, IvezicEtAl2014}), there is no reason why the edge cannot be placed at any point between the photons. In \S \ref{sec:errors} we describe a systematic bias affecting the ``half way'' placement and suggest an adjustment that removes this bias almost entirely. We also consider the case of a continuously varying source intensity, such as those caused by an extended emitting spot moving into or out of eclipse.

Secondly, \emph{XMM-Newton} observations are frequently affected by soft proton flares originating in Earth's magnetosphere that contribute a substantial, and often rapidly variable, background \citep{LumbEtAl2002} that needs to be removed to recover the true variability of the source. Our data therefore consists of photon lists: one extracted from a region of sky surrounding the source and containing both source and background photons, and one taken from a source-free region of sky containing background only. It is not obvious how to perform background subtraction using the Bayesian Blocks algorithm. Removing individual source region photons according to the background rate has been suggested (e.g., \citealt{StelzerEtAl2007}), but we would prefer not to discard any data. An attempt at weighting individual photons in a manner very similar to the ones developed in this paper has been recently applied successfully by \cite{MossouxEtAl2015}. In \S \ref{sec:simulated_data} we generate theoretical source and background light curves, containing transient events occurring at known times. In \S \ref{sec:tests} we propose several modifications to the Bayesian Blocks algorithm to allow for background subtraction, and evaluate them for their ability to recover the transients in the model light curve in the presence of a strong, rapidly varying, background.

\section{ANALYSIS OF ERRORS}
\label{sec:errors}

In this paper we use the geometric prior suggested by \cite{ScargleEtAl2013}. This prior contains one adjustable parameter $p_0$, reflecting the balance between suppressing spurious change points and retaining genuine ones. Throughout this paper we have taken $p_0=0.01$, which selects against false positives at the 99\% level.

\begin{figure}
 \includegraphics{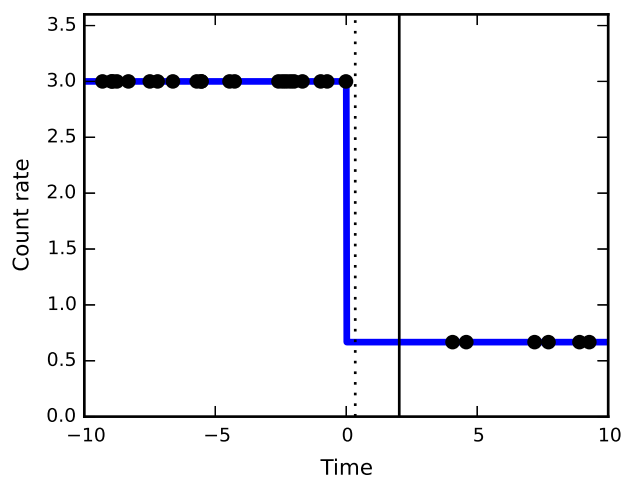}
 \caption{Illustration of the change point bias toward blocks with lower count rate. The two count rates, 3 and 2/3, are shown as the blue line, with events placed according to a Poisson process shown as circles. The half-way point between the two terminal photons is shown as a solid vertical line, as the change point placed according to Equation \ref{eqn:cp_bias_correct} is shown as a vertical dotted line. It is clear that the ``half-way'' change point method is biased towards blocks with lower count rate but that the adjusted change point location gives a better estimate of the true location of the change point.}
 \label{fig:cp_bias_schematic}
\end{figure}

Most implementations of the Bayesian Blocks algorithm place the cell boundaries at the midpoint between two successive events (e.g., \citealt{Scargle1998, ScargleEtAl2013, IvezicEtAl2014}). This will bias the location of change points slightly towards blocks of low count rate. Consider a Poisson process that has rate $r_0$ for times $t<0$ and rate $r_1$ for $t>0$, with $r_1 < r_0$. Suppose the last event of the first segment and the first event of the second segment occur at times $t_0$ and $t_1$ respectively. Since they are both Poisson processes, $t_0$ and $t_1$ will have mean values of $-1/2r_0$ and $1/2r_1$, and the Bayesian Blocks change point, since it falls halfway between the two, will have mean value $(r_0-r_1)/(2r_0r_1)$. This is positive, so the location of the change point is biased toward the block with the lower count rate. Figure \ref{fig:cp_bias_schematic} illustrates this bias. For applications such as eclipse timings in the study of cataclysmic variables, this bias can be a severe drawback, as the count rate during the eclipse is likely to be very low. The effect of this bias will be to cause the eclipse to appear shorter than it really is. Furthermore, if the pre- and post-eclipse blocks are different in count rate, the eclipse midpoint may be shifted.

It is possible to improve the positioning of the change points, as follows: suppose the Bayesian Blocks algorithm has found a block of high count rate beginning at $t_{s}$, containing $n_0$ events, concluding with an event recorded at time $t_0$, followed by a block of lower count rate. We want to find the optimum location of the change point $t_{cp}$. Since we are assuming a Poisson process, $t_0$ is exponentially distributed with standard deviation $1/r_0$ where $r_0$ is the count rate in the block. It follows that, on average, 
\begin{equation}
t_{cp}=t_0+\dfrac{1}{2r_0}.
\end{equation}
We also have
\begin{equation}
r_0 = \dfrac{n_0}{t_{cp}-t_{s}},
\end{equation}
and these can be solved simultaneously to give
\begin{equation}
 t_{cp}=\dfrac{2n_0 t_0-t_{s}}{2n_0-1}.
 \label{eqn:cp_bias_correct}
\end{equation}
The case of a rising count rate is very similar.

We tested this new adjustment with a simulated event series consisting of 100 events with count rate 3, up to $t=0$, followed by another 100 events with count rate $2/3$, and found the change point according to both the ``half way'' method and the one given by Equation \ref{eqn:cp_bias_correct}. We repeated this test for 50,000 realizations of the series. The count rates here are dimensionless because the unmodified Bayesian Blocks algorithm is invariant under changes in the unit of time. The mean location of the recovered change point using the ``halfway'' change point location was $t_{cp}=1.36 \pm 1.83$, comparable to the average spacing of photons in the lower count rate segment and indicating a substantial bias toward the lower count rate. For the new method, the mean location was $t_{cp}=0.461 \pm 1.58$, an improvement by a factor of three. The small bias that remains is entirely due to the algorithm occasionally mistaking the first event of the second segment for an event in the first segment, if it happens to be placed close to the true change in count rate. The converse, mistaking an event in the higher count rate segment for one in the lower count rate segment, is also possible but much less likely since this requires several photons in the higher count rate segment to be badly placed.

We repeated the experiment, discarding all trials with misidentified photons, and found that the mean change point locations for the ``half way'' and adjusted change point locations were $0.66\pm0.45$ and $-0.044\pm 0.20$ respectively. The adjusted change point method suffers from essentially no bias, except for the uncertainty in determining which photons belong to which blocks.

\begin{figure}
     \includegraphics{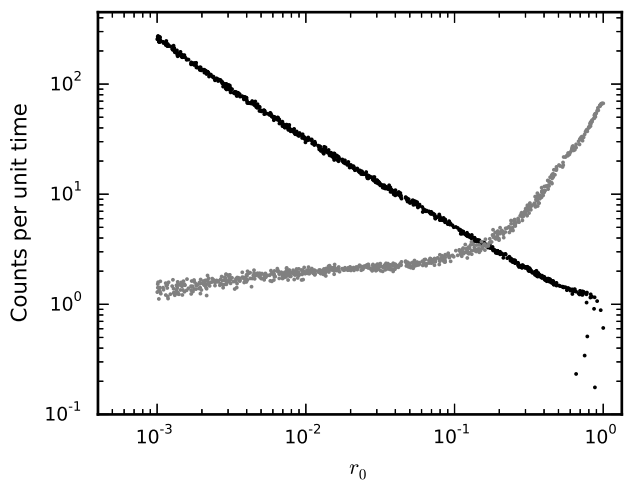}
     \caption{ Reduction in bias by employing the change point adjustment (black points), and remaining uncertainty in change point location (grey points), plotted against the lower of the two count rates. The improvement is larger than the remaining uncertainty for $r_0<0.15$ }
     \label{fig:bias2}
  \end{figure}

  Adjusting the location of change points by this method will only be useful if the uncertainty in their location is small compared to the reduction in bias. To investigate under what circumstances the adjustment is useful we performed the following tests. We produced many segmented light curves as above, consisting of 100 events with count rate $r_1=1$, followed by 100 events of count rate $r_0$ between 0.001 and 1. The division between the two count rates is at $t=0$. We produced 1,000 realizations of each of these light curves and performed the Bayesian Blocks algorithm on them, with and without the new change point adjustment, repeating trials where no change point was found at all. Then we compared the improvement in bias achieved (i.e., the difference of their absolute magnitudes) and compared it to the remaining uncertainty. The results are shown in Figure \ref{fig:bias2}.

  The change point adjustment provides an improvement in bias greater than the remaining uncertainty for $r_0$ smaller than about 0.15 but no useful improvement for higher $r_0$. This behaviour can be understood by considering the sources of uncertainty and bias for the two methods. For the unadjusted method, both bias and uncertainty are dominated by the placement of individual photons in the lower count rate segment. A misidentified photon contributes relatively little to the bias because, to be misidentified, it must be placed within about $1/r_1$ of the change in count rate, and this is smaller than $1/r_0$. For the adjusted method the bias consists almost entirely of misidentified photons, and the uncertainty is partially due to misidentified photons and partially due to the placement of individual photons in the $r_1$ segment. It follows that the bias and uncertainty in the unadjusted method decreases with increasing $r_0$, because $1/r_0$ decreases, and both the bias and uncertainty in the adjusted method increase because misidentified photons become more and more likely. 
  
The adjustment of change points according to Equation \ref{eqn:cp_bias_correct} is therefore most useful when one of the blocks is less than 15\% as intense as the other. For the remainder of the paper, we will be using this method to determine the location of the change points.

\subsection{Continuously varying count rate}

In studies of eclipsing CVs, it is useful to determine or constrain the size of the emitting accretion spot on the white dwarf. As the spot passes behind the secondary star, its observed flux declines steadily to zero. If this decline is gradual enough, the Bayesian Blocks algorithm will produce blocks of intermediate count rate during the ingress or egress. If the decline is too rapid, the algorithm will not resolve it and produce only an instantaneous change in count rate; the failure can still be used to constrain the size of the emitting region.

\begin{figure}
 \includegraphics[width=80.005mm]{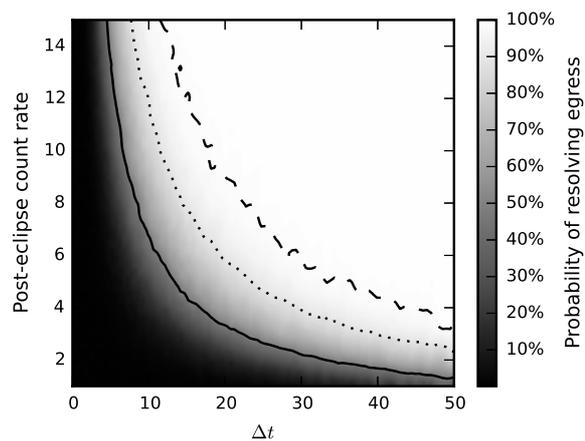}
 \caption{ Probability of resolving an eclipse egress with post-eclipse count rate $R$ and eclipse egress duration $\Delta t$. Contours are shown at the 50\%, 90\%, and 99\% significance level (solid, dotted, and dashed lines respectively). The probability of finding more than one change point during egress increases with increasing duration and post-eclipse count rate. }
 \label{fig:slopedetect}
 
\end{figure}

We investigated this issue by simulating many hypothetical eclipse egresses, with a very low eclipse count rate of $10^{-3}$ and post-eclipse count rates $R$ between 1 and 15. The egress itself was modelled as a linearly rising count rate with duration $\Delta t$ between 0 and 50 and the fiducial time $t=0$ was placed at the midpoint of the egress. For each set of $R$ and $\Delta t$ we made 1,000 realizations and found the Bayesian Blocks change points. Figure \ref{fig:slopedetect} shows the probability that the egress will be resolved, that is, that the Bayes Blocks algorithm finds more than one change point in the egress. It is clear that this probability increases with increasing egress length, and with increasing post-eclipse count rate, as one would expect.

The usefulness of this analysis to eclipse timing is clear. If the ingress or egress is not resolved, one can place constraints on the size of the emitting spot by finding the smallest $\Delta t$ for which that ingress or egress would have been resolved. For instance, if an eclipse egress in a real observation is not resolved and the post-eclipse count rate is 6 photons/s, it can be seen from Figure \ref{fig:slopedetect} that the egress has a shorter duration than 30~s with approximately 99\% confidence.

\begin{figure}
 \includegraphics[width=80.005mm]{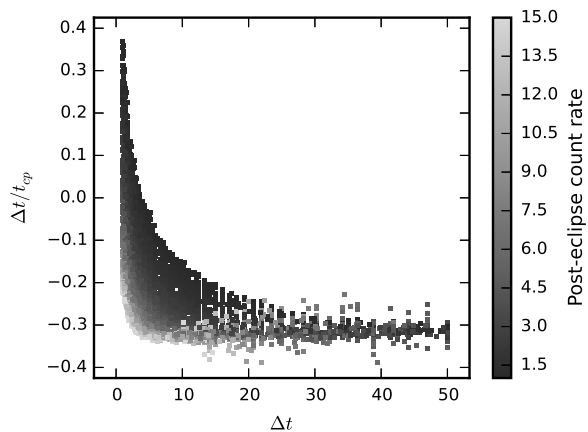}
 \caption{ Bias in the eclipse egress change point location as a fraction of egress duration, for simulations in which the egress was not resolved. Colors indicate the post-eclipse count rate.}
 \label{fig:egress_bias}
\end{figure}

For simulations where the egress was not resolved, that is, where only one change point was found for the egress, we found the mean position of that change point. The results are shown in Figure \ref{fig:egress_bias}. Three behaviours are evident in this Figure. For short eclipse egresses, the situation closely resembles an instantaneous jump in count rate and the bias on the position of the change point is accordingly scattered around zero, with the scatter appearing large on the Figure because $\Delta t$ is small. For intermediate durations, where the duration is long compared to the photon separation in the post-eclipse block, the change point is triggered earlier than the half-way point of the egress. There is a bias of about 25 to 33\%. Finally, when the post-eclipse count rate is high and the egress is long, then the algorithm never fails to resolve it, as indicated by the absence of such points in Figure \ref{fig:egress_bias}.

\section{SIMULATED DATA FOR BACKGROUND SUBTRACTION}
\label{sec:simulated_data}

In this section we devise and apply several methods for performing background subtraction using a Bayesian Blocks approach. We have tested the various methods in this paper on a hypothetical \emph{XMM-Newton} observation of a binary system exhibiting transients in the form of eclipses and bursts. The observation of the source is affected by a soft proton flare, contributing a substantial background of photons that increases in magnitude over the observation, and is highly time-variable. 

We generated two lists of photon arrival times. The first is assumed to be taken from a region surrounding the source, containing photons from the source itself and from the soft proton flare superimposed upon it. The second list contains only flare photons, taken from a region 4.123 times larger than the source extraction region. This number was arbitrarily chosen. Thus, photons from the background extraction region will be given a weight of $W_B=1/4.123$ compared to source region photons. We will attempt to recover the source light curve by subtracting the soft proton flare from the total light curve. The ability of our background subtraction methods to recover the timings of the transient events is our measure of the suitability of the background subtraction methods. We summarize the results in Table \ref{tab:transient_times} and Figure \ref{fig:deltas}.

The source and flare light curves are described in \S \ref{sec:source_light_curve} and \S \ref{sec:background_light_curve} respectively.

\subsection{Source light curve }
\label{sec:source_light_curve}
We have taken the behaviour of the LMXB EXO~0748$-$676 as a model for our simulated source light curve. This system exhibits eclipses of 8.3 minute duration, with an orbital period of 3.82 hours \citep{ParmarEtAl1986}, during which the X-ray flux from the source drops to zero. It also undergoes type-I X-ray bursts, during which the X-ray flux increases almost instantaneously to many times its pre-burst level \citep{GottwaldEtAl1986}. When neither eclipses nor bursts are present, the source emits with a count rate of about 3~counts/s in \emph{XMM-Newton} observations \citep{HomanEtAl2003}.

\begin{figure}
 \includegraphics{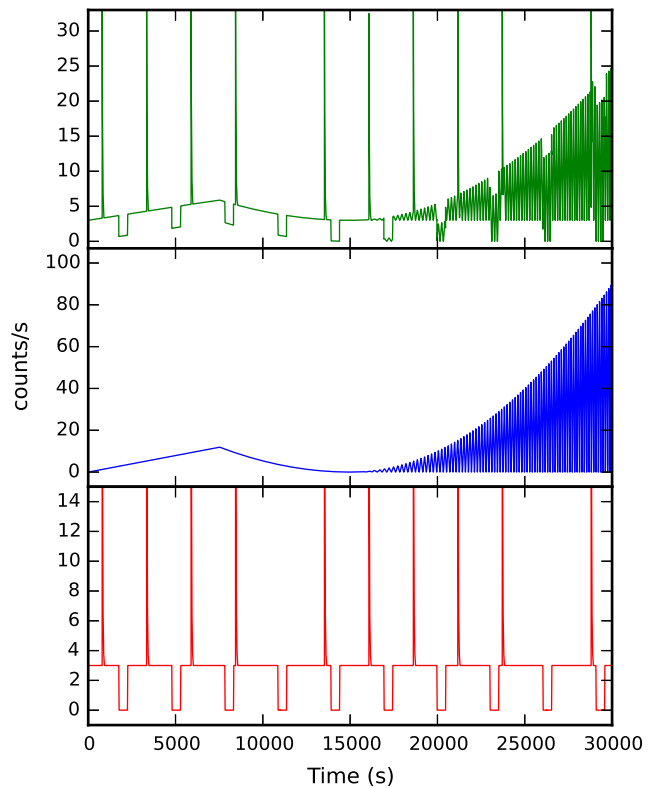}
 \caption{ Total source light curve (top panel), background (middle panel), and zero-background source light curve (bottom panel). The high intensity of the background signal is due to the higher exposure area (4.123 times greater than the source area).}
 \label{fig:functions}
\end{figure}

The simulated source light curve consists of a constant persistent intensity of 3~counts/s. Superimposed upon this are eclipses of duration 498~s, recurring with a faster orbital period of 3,035~s, and bursts recurring with a period of 2545.6~s. The bursts have peak intensity of 30~counts/s and exponential decay time of 24~s. The transient events in the light curve are therefore (see Figure \ref{fig:functions}) ten bursts, ten eclipse ingresses, and ten eclipse egresses. All transients occur instantaneously, and two bursts are ``missing'' due to falling inside an eclipse. The times of these events are given in Table \ref{tab:transient_times}. There are 154,992 photons in this series.

\subsection{Background light curve}
\label{sec:background_light_curve}

The background consists of three segments. First is a linear rise from a count rate of 0 to 24, to $t=7500$, followed by a quadratic fall back to zero. These two segments are intended to investigate whether the presence of a rising or falling background contribution biases the timings of transients. Beginning at 15,000~s is an oscillatory signal with increasing amplitude and frequency, designed to approximate the behaviour of the soft proton flares that affect observations by X-ray instruments such as \emph{XMM-Newton} and \emph{Chandra} \citep{LumbEtAl2002}. Its count rate is given by the formula

\begin{equation}
 R(t) = 2.0\times 10^{-7} t_a^2\left[1+\sin\left(\frac{1}{5000}t_a^{1.25}\log_e t_a\right)\right]\text{ photons/s,}
\end{equation}
where $t_a=t-15000$. There are 300,725 photons in this series.

\section{BACKGROUND SUBTRACTION}
\label{sec:tests}

\subsection{Constant Cadence}

\begin{figure}
 \includegraphics{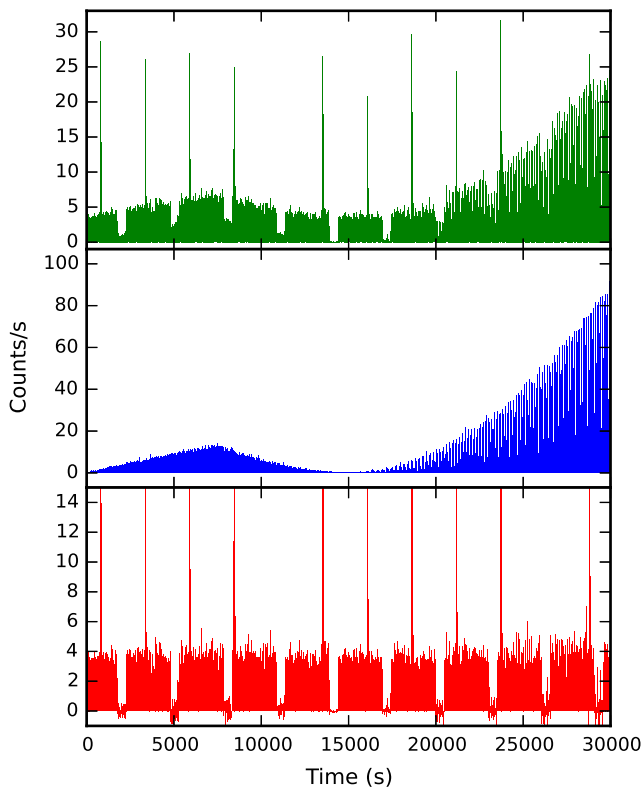}
 \caption{ Source, background, and background subtracted light curves (top, middle and bottom panels respectively) accumulated into bins of predetermined width (10~s) and position. This method is effective at removing the background but its ability to time transient events is limited by the previously set bin size.}
 \label{fig:histos}
\end{figure}

Before analyzing various implementations of the Bayesian Blocks algorithm, we study a case in which the locations of the change points are \emph{not} optimally determined by the data, as a comparison. The simplest method is to simply group the source and background photons into equally spaced bins whose width and location are fixed beforehand, then subtracting the latter from the former to obtain the background subtracted light curve. In Figure \ref{fig:histos} we show the two photon series, and their difference, counted in bins of 10~s width.

A property of the constant cadence method is that signals with more rapid variability than the cadence will tend to be averaged out. This property makes it useful for subtracting a rapidly varying background. However, fast transients like burst rises and eclipse in- and egress (instantaneous in our simulated data) will not be timed accurately unless they fortuitously happen to fall on or near a bin boundary. An instantaneous transient event falling well inside a bin will not appear instantaneous because it will produce one bin of intermediate count rate, giving the false impression of a gradual change; this phenomenon is clearly seen in panel \emph{b} of Figure \ref{fig:insets}. In this regime, the uncertainty in the timing of the transient event will be less than $\Delta t/2$, where $\Delta t$ is the width of the bins.

A further disadvantage of the constant cadence method is that it is necessary to select the width of the bins beforehand. Various estimates for the optimum bin width have been proposed (see \citealt{BirgeRozenholc2006} for a discussion), but they often suggest too few bins to perform timing analyses. The commonly used rule of \cite{Sturges1926}, taking $1+\log_2N$ intervals, gives only 18 or 19 bins for the 154,992 source photons. Similarly, the rules of \cite{Scott1979} and \cite{FreedmanDiaconis1981} give bin widths of hundreds of seconds.

\subsection{Bayesian Blocks- Direct Subtraction}

\begin{figure}
 \includegraphics{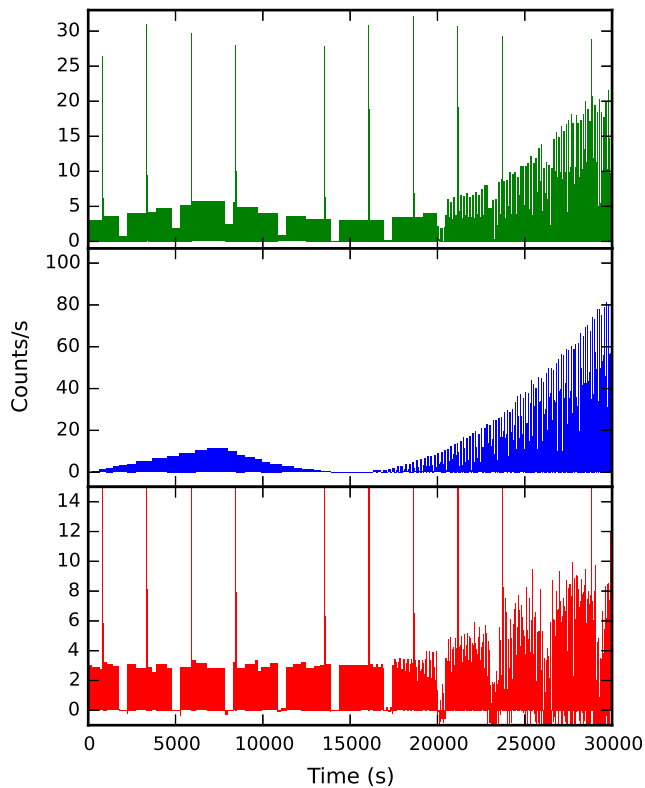}
 \caption{ Direct subtraction of the Bayesian Blocks representation of the background from that of the source. This method produces one change point for each change point in the two Bayesian Blocks representations from which it is derived.}
 \label{fig:direct_subtract}
 
\end{figure}

The most obvious way to perform background subtraction is to generate Bayesian Blocks light curves for the source and background regions and subtracting the latter from the former, after normalizing the background series to account for the larger size of the extraction region. The results of this test are shown in Figure \ref{fig:direct_subtract} and panel \emph{c} of Figure \ref{fig:insets}. The resulting background-corrected light curve will contain a block boundary for every change point in both of the source and background Bayesian Blocks representations, and many of these will be spurious. Subtracting the background Bayesian Blocks light curve from the source light curve succeeds in removing much of the background. However, the resulting light curve is very noisy, particularly towards the end of the observation. 

\cite{ScargleEtAl2013} provide a method for combining multiple data series in such a way that the change points in each series line up. Such a process is useful for concurrent observations of the same transients by different instruments, and will not produce superfluous change points when they are added or subtracted, but is not suitable for data series such as ours, which contain completely different transients.

If we do not wish to produce a Bayesian Blocks representation of the background, the background can be subtracted from the source Bayesian Blocks representation on the level of blocks, cells, or individual photons. These three approaches are described in \S \ref{sec:weightedblocks}, \ref{sec:weightedcells}, and \ref{sec:weightedphotons}.

\subsection{Bayesian Blocks- Weighted Blocks}
\label{sec:weightedblocks}

\begin{figure}
 \includegraphics{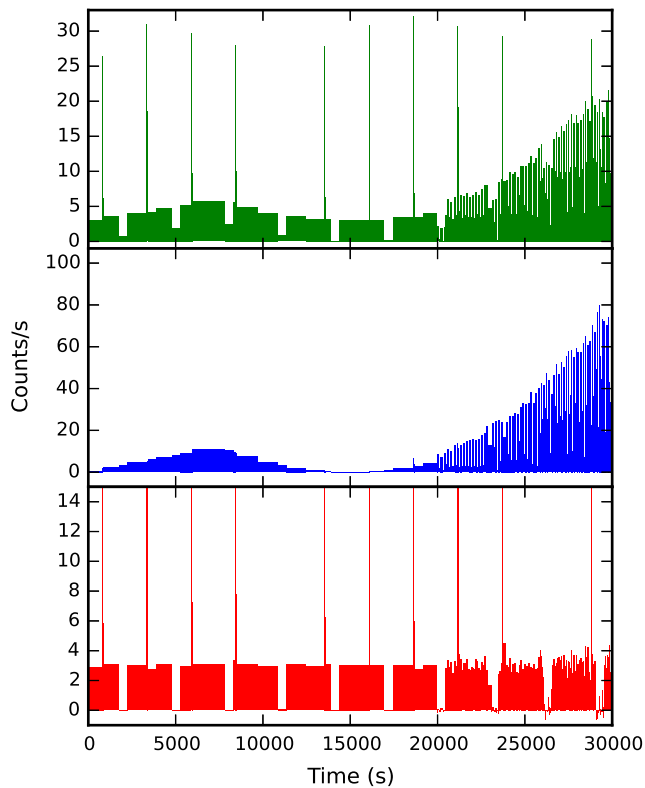}
 \caption{Background corrected light curve using the ``weighted blocks'' method. The background subtraction appears to be very effective, producing almost no noise even at the end of the observation. The eighth and ninth eclipses appear to have shallowly sloping edges, probably due to the presence of pulses in the background.}
 \label{fig:weighted_blocks}
 
\end{figure}

This method is suggested by the observation that photons from the background region are more numerous, but individually carry little weight, compared to source photons. Here we find the Bayesian Blocks change points of the source photon list, and subtract from the source photon count rate the exposure area weighted count rate of the background photons falling into that block. That is,
\begin{equation}
 CR=\dfrac{n_S - W_Bn_B}{L},
\end{equation}
where $CR$ is the background subtracted count rate, $n_S$ and $n_B$ are the numbers of source region and background region photons respectively, and $L$ is the length of the block. The results are shown in Figure \ref{fig:weighted_blocks} and in panel \emph{d} of Figure \ref{fig:insets}.

Although this method does not produce as many spurious change points as the previous one, it is clear that it cannot give the locations of the transients as accurately, because it contains only the change points of the source series and these are potentially offset from their true locations by the variable background, as can clearly be seen in Figure \ref{fig:insets}.

\subsection{Bayesian Blocks- Weighted Cells}
\label{sec:weightedcells}

In this approach we subtract the weighted background photons from each cell before the change points are found. Thus, each cell contains not one photon, but $1-n_BW_B$ photons, where $n_B$ is the number of background photons occurring within that cell. As the fitness function of each potential block (Equation 19 of \citealt{ScargleEtAl2013}) involves the logarithm of the counts in blocks, we must find a way of dealing with cells and blocks with nonpositive count rates. If the count rate is not positive we set the effective count rate in the logarithm to a small positive number $s_\text{min}$ (we have taken $s_\text{min}=1.0\times 10^{-4}$). The cells themselves are defined by the source photons, and therefore each contains at least one positively weighted photon, hopefully reducing the number of times we need to resort to this arbitrary countermeasure. If a block edge separates two blocks of apparently negative count rate, we do not adjust its location according to Equation \ref{eqn:cp_bias_correct} because it is unclear what the spacing between two hypothetical photons in such a block should be.

\begin{figure}
 \includegraphics[width=80mm]{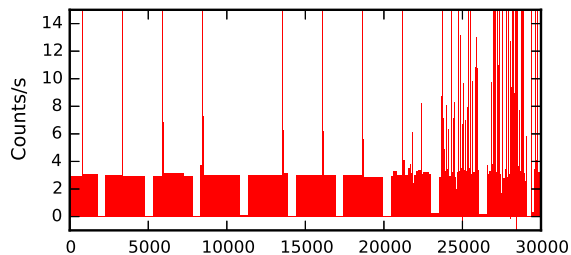}
 \caption{Background corrected light curve using the ``weighted cells'' method. There is considerable noise near the end of the observation, but all the transients appear sharply defined.}
 \label{fig:weightedcells}
\end{figure}

The results of this trial are shown in Figure \ref{fig:weightedcells} and panel \emph{e} of Figure \ref{fig:insets}. The ``weighted cells'' method performs well throughout most of the observation, but towards the end it produces many short spurious blocks with implausibly high count rates. It also badly misses some eclipse timings (see panel \emph{e} of Figure \ref{fig:insets}, and Table \ref{tab:transient_times}).

\subsection{Bayesian Blocks- Weighted Photons}
\label{sec:weightedphotons}
This approach is similar to the previous one, except that we no longer define the cell edges by source photons only. Instead, we combine both photon lists into one, and each cell has a weight of either 1 or $-W_B$, depending on whether it contains a source or a background photon. As in the previous method, we do not adjust the location of the change point between two blocks with negative count rates.

\begin{figure}
 \includegraphics[width=80mm]{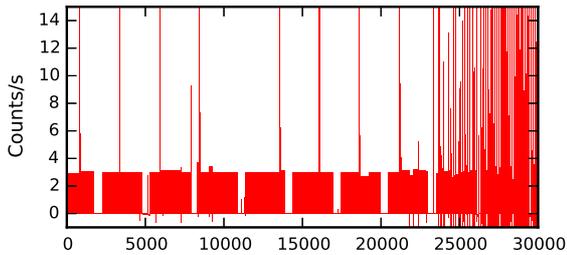}
 \caption{Background corrected light curve using the ``weighted photons'' method. Noise, in the form of short blocks with implausibly high or low count rates, is more prominent in this method, but the transients are sharply defined.}
 \label{fig:weighted_photons}
\end{figure}

The problem with negative count rates is more prominent now, because there are many cells that contain a background photon and therefore have negative count rates. However, the placement of change points is potentially finer. The results are shown in Figure \ref{fig:weighted_photons} and panel \emph{f} of Figure \ref{fig:insets}. There are many short blocks with absurdly high or low count rates near the end of the simulated observation, and even a few nonsense blocks near the beginning. The timing properties of this approach are excellent nonetheless.

\subsection{Bayesian Blocks- Iterated Bayesian Blocks}

\begin{figure}
 \includegraphics{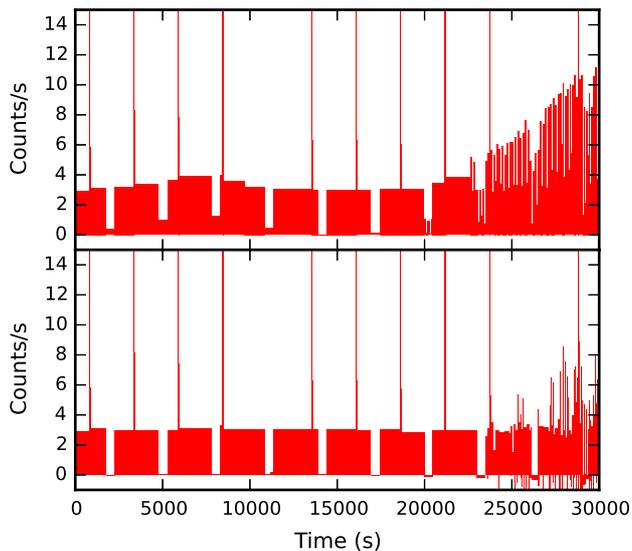}
 \caption{The method described in \cite{MossouxEtAl2015}. The top panel shows the original photon weighting and the bottom panel shows the alternative photon weighting. The original weighting does not actually subtract the background, but only attempts to locate the transients, which explains why the eclipses do not have count rates near zero. Both methods are only noisy near the end of the observation.}
 \label{fig:m15}
\end{figure}

\begin{figure}
 \includegraphics{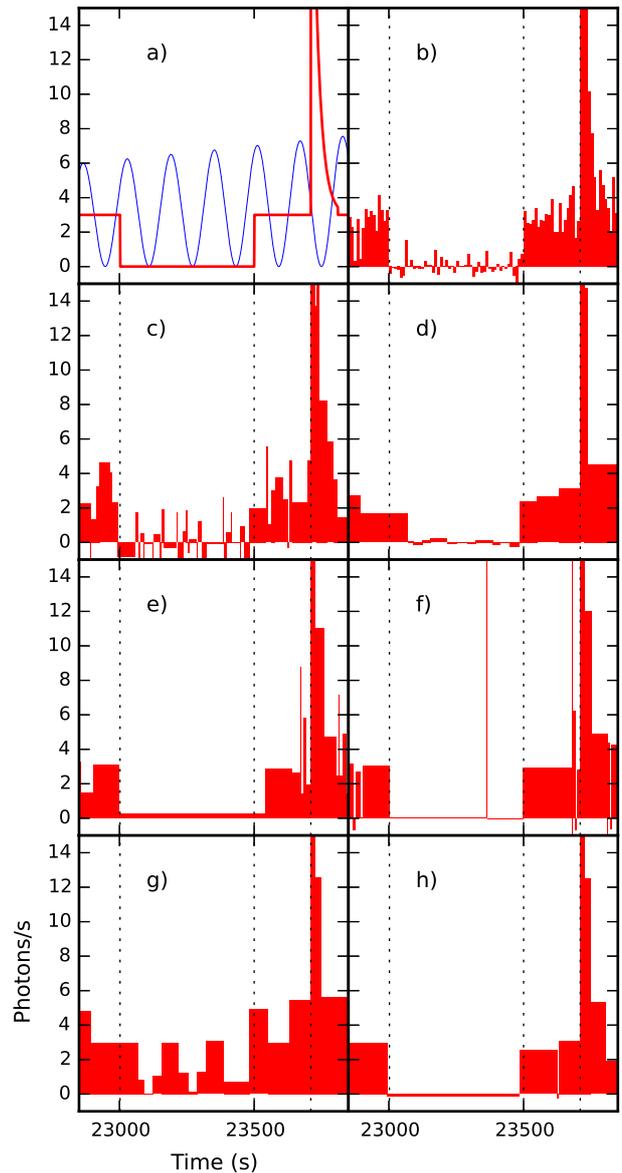}
 \caption{The region around eclipse \#8 and burst \#9. a) true source light curve (red) and background (blue), b) constant cadence, c) direct subtraction, d) weighted blocks, e) weighted cells, f) weighted photons, g) iterated Bayesian Blocks with original weighting, h) iterated Bayesian Blocks with alternative weighting. The true locations of the transients are indicated by dotted vertical lines.  All other methods locate the burst accurately but many fail to accurately time the eclipse ingress, which coincides with a pulse of the variable background. The ``weighted photons'' approach performs very well in locating all transients.}
 \label{fig:insets}
\end{figure}

This approach is the one developed and successfully applied to X-ray activity of Sgr A* by \cite{MossouxEtAl2015}. It is effectively a hybrid of the weighted cells and direct subtraction approaches. Here, a Bayesian Blocks representation is produced for the background and source light curves to obtain count rates for the source and background regions. The count rates are then used to calculate appropriate weights for photons in the source region and the algorithm is then run a third time. A photon falling within a block of source region count rate $C_S$ and background region $C_B$ is given a weight of $w=C_S/(C_S+C_B)$, where $C_B$ is positive and corrected for exposure area. It is clear to see that weighting the photons in this way does not actually subtract the background, but is intended only to find the locations of change points. For this reason we also test this recipe with an alternate weighting, $w=1-C_B/C_S$. The results are shown in Figure \ref{fig:m15}, and panels \emph{g} and \emph{h} of Figure \ref{fig:insets}. 

\begin{table*}
\begin{tabular}{lrrrrrrrr}
Transient & Time (s) & $\Delta t_{b}$ & $\Delta t_{c}$ & $\Delta t_{d}$ & $\Delta t_{e}$ & $\Delta t_{f}$ & $\Delta t_{g}$ & $\Delta t_{h}$\\
\hline
I-1  &  1751.0  &  -1.0  &  0.4  &  0.4  &  0.1  &  0.5  &  1.7  &  1.7\\
I-2  &  4786.9  &  3.1  &  -0.2  &  -0.2  &  0.4  &  0.1  &  0.4  &  0.4\\
I-3  &  7822.7  &  -2.7  &  4.0  &  4.0  &  -3.0  &  3.7  &  4.4  &  4.4\\
I-4  &  10858.6  &  1.4  &  -0.1  &  -0.1  &  -2.3  &  -0.1  &  0.7  &  0.7\\
I-5  &  13894.5  &  -4.5  &  0.1  &  0.1  &  -0.7  &  0.1  &  8.9  &  8.9\\
I-6  &  16930.3  &  -0.3  &  0.1  &  0.1  &  -3.2  &  0.1  &  2.4  &  2.4\\
I-7  &  19966.2  &  3.8  &  -0.2  &  -0.2  &  -0.2  &  0.0  &  6.7  &  6.7\\
I-8  &  23002.1  &  -2.1  &  -8.3  &  68.3  &  -3.4  &  0.5  &  68.9  &  -8.2\\
I-9  &  26037.9  &  2.1  &  -5.0  &  25.2  &  0.8  &  0.5  &  25.0  &  5.5\\
I-10  &  29073.8  &  -3.8  &  1.1  &  1.1  &  -0.6  &  1.2  &  2.1  &  2.1\\
\hline
E-1  &  2249.0  &  1.0  &  -0.0  &  -0.0  &  0.5  &  0.2  &  -1.4  &  -1.4\\
E-2  &  5284.9  &  -4.9  &  0.3  &  0.3  &  0.5  &  -0.3  &  -0.2  &  -0.2\\
E-3  &  8320.7  &  -0.7  &  -1.3  &  -1.3  &  2.0  &  0.0  &  -1.8  &  -1.8\\
E-4  &  11356.6  &  3.4  &  -0.3  &  -0.3  &  0.1  &  -0.0  &  -3.5  &  -3.5\\
E-5  &  14392.5  &  -2.5  &  -0.1  &  -0.1  &  0.7  &  0.9  &  -23.2  &  -23.2\\
E-6  &  17428.3  &  1.7  &  -1.4  &  -1.4  &  0.4  &  0.1  &  -2.5  &  -0.5\\
E-7  &  20464.2  &  -4.2  &  -4.9  &  -4.9  &  -0.1  &  -4.4  &  -5.4  &  -5.4\\
E-8  &  23500.1  &  -0.1  &  -16.4  &  -16.4  &  40.9  &  -1.1  &  -16.8  &  -14.1\\
E-9  &  26535.9  &  4.1  &  16.2  &  44.0  &  35.2  &  -0.3  &  44.4  &  16.0\\
E-10  &  29571.8  &  -1.8  &  -0.3  &  10.7  &  15.5  &  -0.7  &  21.5  &  -0.3\\
\hline
B-1  &  800.0  &  0.0  &  -0.0  &  -0.0  &  0.0  &  0.1  &  -0.3  &  -0.3\\
B-2  &  3345.6  &  4.4  &  -0.0  &  -0.0  &  0.1  &  0.1  &  -0.2  &  -0.2\\
B-3  &  5891.2  &  -1.2  &  -0.0  &  -0.0  &  0.1  &  0.0  &  -0.3  &  -0.3\\
B-4  &  8436.8  &  3.2  &  -0.1  &  -0.1  &  -0.6  &  -0.6  &  -0.2  &  -0.2\\
B-5  &  13527.9  &  2.1  &  -0.1  &  -0.1  &  0.0  &  0.0  &  -0.4  &  -0.4\\
B-6  &  16073.5  &  -3.5  &  0.0  &  0.0  &  0.2  &  0.2  &  -0.2  &  -0.2\\
B-7  &  18619.1  &  0.9  &  -0.0  &  -0.0  &  0.0  &  0.0  &  -0.2  &  -0.2\\
B-8  &  21164.7  &  -4.7  &  0.0  &  0.0  &  0.1  &  0.1  &  -0.0  &  -0.4\\
B-9  &  23710.3  &  -0.3  &  -0.0  &  -0.0  &  0.1  &  0.0  &  -0.1  &  -0.1\\
B-10  &  28801.4  &  -1.4  &  0.1  &  0.1  &  -0.9  &  0.1  &  -0.1  &  -0.1\\
\hline
\end{tabular}
\caption{Timing accuracy of the background subtraction methods for the thirty simulated transients. The transients are grouped by type rather than chronological order, with I, E, B referring to ingreses, egresses, and bursts. The columns $\Delta t_b$ through $\Delta t_f$ correspond to the error in the timings: b= constant cadence, c=direct subtraction, d=weighted blocks, e=weighted cells, f=weighted photons, g=Iterated Bayesian Blocks method with original weighting, h=Iterated Bayesian Blocks method with with alternate weighting.}
\label{tab:transient_times}
\end{table*}

\subsection{Invariance to time units}

The unmodified Bayesian Blocks algorithm produces the same change points regardless of the choice of time unit. We have investigated whether this desirable behaviour is preserved for the various modifications described above. We took the same photon series and divided the arrival times by 3600 to express the arrival times in hours rather than seconds. Naturally the constant cadence, direct subtraction, and weighted blocks methods are unchanged under this transformation, but all other methods produced many more change points than previously. The ``weighted photons'' method, for example, produced 16,080 change points for the time-scaled photon series, compared to 971 for the original series. This behaviour is probably due to our safeguard against negative block count rates. Since we have achieved tolerably good results with count rates of 1 to 10, but higher effective count rates produce very many spurious change points, it may be generally desirable to scale the time unit. We have also found that the number of change points produced is fairly insensitive to the choice of $s_\text{min}$.

It would be desirable to have a method for dealing with blocks of negative count rate that has a better theoretical justification. Similar questions have previously been considered. \cite{Loredo1992}\footnote{ A longer and more comprehensive version of this book chapter is available at \url{http://www.astro.cornell.edu/staff/loredo/bayes/tjl.html}} gives a posterior probability distribution for the source count rate, independent of the count rate of the background, which could potentially be used to ensure positive count rates in all potential blocks. A method developed by \cite{Zech1989} gives the probability of the source having a positive count rate $s$ given the number of total (source + background) photons detected, and the background count rate. This can be used to estimate the largest plausible source count rate. However, it is not immediately obvious how to incorporate either of these procedures into the fitness function of Equation 19 in \cite{ScargleEtAl2013}. Both procedures also require summations over all the photons in a block, increasing the running time of the algorithm from $\mathcal{O}(n^2)$ to $\mathcal{O}(n^3)$ or worse.

\subsection{Background exposure area}

We investigated the effect of changing the exposure area of the background by repeating the ``weighted photons'' method on two new simulated observations, using background extraction areas of 2.062 and 8.246 as large as the source extraction region, that is, half and double the exposure area we have been using up to now. The background photons were weighted according to these new exposure areas. 

For the three background exposure areas, we found that the smallest one produced 1,302 change points, the middle one produced 971 change points, and the largest background exposure area produced 742. The accuracy in timing the thirty transient events was not adversely affected by increasing or decreasing the background exposure area, but the larger this area, the less spurious change points were produced.

\section{DISCUSSION AND CONCLUSION}

We have investigated the ability of the Bayesian Blocks algorithm to accurately and precisely time transient events. With an adjustment that moves the change point toward the block of higher count rate, the algorithm is able to determine the locations of instantaneous increases or decreases in count rate with essentially no systematic bias and with uncertainties similar in size to the interval between individual photons. 

When the change in count rate is not instantaneous, we have characterised the ability of the algorithm to resolve the period of varying count rate and, if it cannot be resolved, found that the change point is placed near the block of lower count rate. These observations have important implications to applications such as measuring the times and durations of eclipses.

\begin{figure}[h!]
 \includegraphics{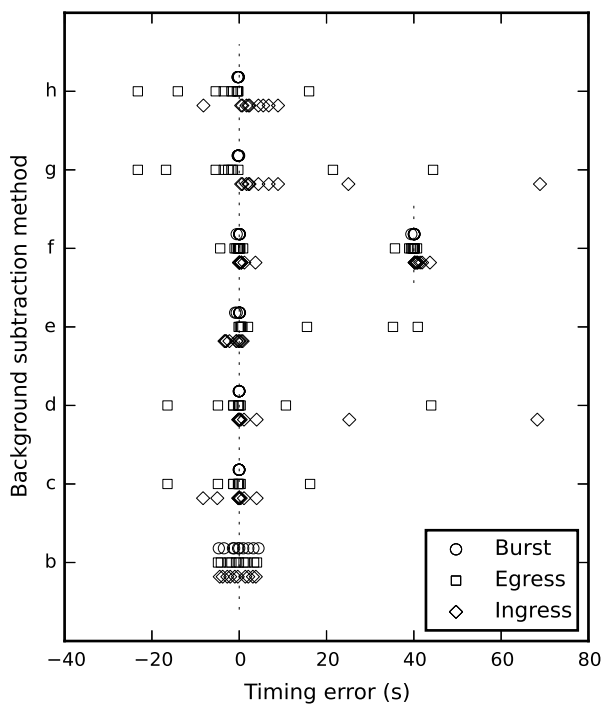}
 \caption{ Timing errors of the thirty transient events, for the different background subtraction methods. The letter identifiers are the same as in Table \ref{tab:transient_times} and Figure \ref{fig:insets}: b) constant cadence, c) direct subtraction, d) weighted blocks, e) weighted cells, f) weighted photons, g) iterated Bayesian Blocks with original weighting, h) iterated Bayesian Blocks with alternative weighting. The transient events have been divided by type. The constant cadence method finds all types of transients with precision limited by the bin size. All other methods perform well for bursts, but only the ``weighted photons'' (f) approach is reliably accurate at timing eclipse ingresses and egresses. The offset points in row (f) show the timing errors for this method if the locations of the change points are not adjusted according to Equation \ref{eqn:cp_bias_correct}. Dotted vertical lines indicate a timing error of zero.}
 \label{fig:deltas}
\end{figure}

We have incorporated background subtraction into the Bayesian Blocks algorithm in several different ways, and tested the alternatives against simulated source and background signals containing bursts and eclipses of known position. Even when the observation is dominated by large, rapidly varying, background contamination it is possible to recover the transient times with excellent accuracy. In Table \ref{tab:transient_times} we show the time locations and nature (e.g., I-3 is the third eclipse ingress) of the transients, as well as the error (the distance to the nearest bin edge found by each of the background subtraction methods). The results are also shown graphically in Figure \ref{fig:deltas}. Note that, because we know beforehand where the transients are, we can unambiguously identify the nearest bin edge. In real applications it may be difficult to determine which bin edge actually marks the beginning of the transient, so it is likely that methods producing many needless bin edges or change points will actually perform worse on real data than in this paper. Visual inspection of the results is always called for.

The ``weighted photons'' approach (\S \ref{sec:weightedphotons}) is clearly best for background subtraction. As shown in Table \ref{tab:transient_times}, it correctly recovers all transients to within a few seconds and most to within half of a second.  However, it is prone to producing occasional blocks of very short lengths and implausibly high count rates (see Figures \ref{fig:weighted_photons} and \ref{fig:insets}), but if this drawback can be tolerated then the ``weighted photons'' adaption to the Bayesian Blocks algorithm is well suited for transient timing. All alternatives perform well in locating the bursts, but only the ``weighted photons'' approach reliably recovers the eclipse ingresses and egresses. The reason for this can be seen in Figure \ref{fig:insets}; a pulse of background contamination coincides with an eclipse ingress, and this causes the ``weighted' blocks'' and unmodified iterated Bayesian Blocks methods to miss the ingress entirely, and most of the other alternatives can recover it only approximately.

The effect of adjusting the change points according to Equation \ref{eqn:cp_bias_correct} is also shown in Figure \ref{fig:deltas}. The offset points for the ``weighted photons'' row show the effect of not performing this adjustment. There is little visible difference, but the timings of some of the ingresses are visibly improved by performing the change point adjustment.

We included a slowly rising and falling background in the simulated observation to determine if this causes a systematic offset in the timings but, from Table \ref{tab:transient_times}, there is no indication of such an effect in any of the alternatives we considered.

The modifications to the Bayesian Blocks algorithms developed in this paper can be generalised to deal with simultaneous observations with different instruments, each with their own source and background extraction regions. One simply assigns every photon from all photon series a weight according to the instrument's extraction area and sensitivity.

The unmodified Bayesian Blocks algorithm is invariant under choice of time unit. That is, it makes no difference to the number and placement of the change points if the photon arrival times are expressed in units of seconds, hours, orbital phase, etc. We have found that this useful property is not preserved for modifications to the Bayesian Blocks algorithm in which cells or blocks can have negative weight, probably due to the way we avoid taking the logarithm of a negative photon count. The higher the count rate, the more spurious change points will be produced. If the unit of time is scaled to achieve typical count rates of 1-10, the results should still be acceptable. 

\begin{acknowledgements}
 This work was supported by the German DLR under contract 50 OR 1405. This research made use of Astropy, a community-developed core Python package for Astronomy (Astropy Collaboration, 2013)\nocite{Astropy2013}. We are grateful to the anonymous referee, whose suggestions significantly improved this work.
\end{acknowledgements}

\bibliographystyle{apj}
\bibliography{bbbib}

\end{document}